\documentclass[intlimits,twoside,a4paper]{article}

\usepackage{amsmath,amssymb}
\usepackage{graphicx, subfigure}

\usepackage[T2A]{fontenc}
\usepackage[cp1251]{inputenc}

\usepackage[eqsecnum]{cmpj2}

\usepackage{textcomp}
\issue{2014}{17}{4}{44701}
\doinumber{10.5488/CMP.17.44701}

\articletype{Rapid Communication}

\title[Longitudinal evolution of magnetization]%
{Longitudinal evolution of magnetization of ferromagnets following
ultrafast demagnetization: Role of finite size and shape of demagnetized region}
\author[I.A. Yastremsky]{I.A. Yastremsky}
\address{
Taras Shevchenko National University of Kyiv, 64 Volodymyrska St., 01601 Kyiv, Ukraine}

\date{Received September 4, 2014, in final form October 8, 2014}
\authorcopyright{I.A. Yastremsky, 2014}

\begin{document}
\maketitle
\begin{abstract}
A dependence of a relaxation rate on the shape of the demagnetized region for a longitudinal evolution of total magnetization to its equilibrium value following the ultrafast demagnetization is demonstrated. This shape-dependence is caused by a motion of the wave front inside the demagnetized region. The contribution of the wave front for spherically symmetric shape of the demagnetized region is up to 3 times and for cylindrically symmetric shape up to 2 times stronger than for one dimensional demagnetized region. This effect can be observed after the demagnetization by a tightly focused femtosecond laser pulse.
\end{abstract}

\section{Introduction}

Recently, in the works where for a tightly focused
(up to 1~\textmu{}m) femtosecond laser pulse it was possible
to observe the real picture of the spatial-temporal evolution under
the action of a solitary pulse, the authors reported on the
development of strongly inhomogeneous states. Along with the
propagation of spin waves far away from the light spot
\cite{Terui12, Dvornik13}, the authors observed the formation of
domains \cite{Jong12} and ring structures localized near the spot
\cite{Dvornik13}. Strongly inhomogeneous distributions of the
magnetization appear in the framework of superdiffusive mechanism of
action of the laser pulse on magnetic metals \cite{Battiato10,
Battiato12} and the existence of these distributions is established for magnetic
heterostructures \cite{Rudolf12, YOI14}. Thus, the analysis of the
picosecond, longitudinal evolution of inhomogeneous, nonequilibrium
distributions of magnetization has become increasingly important.

The analysis of such states can be performed using the Landau-Lifshitz
equation with a relaxation term proposed by Bar'yakhtar
\cite{Bar-jetp84,Bar-ftt,Bar-PhyB}, also  referred to as LLBar equations
\cite{dvornikPRB13}. LLBar equations were derived using general
principles (Onsager's relations, symmetry of the exchange
interaction), and they have no alternatives for the study
of the highly non-linear, nonequilibrium, non-uniform states with strong reduction of magnetization mentioned above.
LLBar equations are well suited for a description of non-uniform
states, such as magnetic solitons \cite{BarIvSobSuk,BarIvSaf} and Bloch
points \cite{GalkinaStef}, and give the explanation
of the reversal effects \cite{MentinkPRL12,Bar+JLett13}. These equations provide an
explanation \cite{YOI14} of recent experiments \cite{Rudolf12} on magnetization recovery in
laser-pumped Ni-Ru-Fe heterostructures, where the importance of the nonlocal character of the magnetization recovery
is established \cite{YOI14}.

A longitudinal relaxation of a total magnetization of ferromagnets following the ultrafast demagnetization within one dimensional (1d) model is determined by two effects: a homogeneous evolution of a
magnetization inside the demagnetized region and the motion of the wave front from
the outside to within the demagnetized region \cite{FNT}. For a
sufficiently large diameter of the demagnetized region, a
homogeneous evolution dominates. However, for a smaller
demagnetized region, a relative contribution of the wave front
increases and for tightly focused femtosecond laser pulse the motion
of the wave front can significantly enhance the relaxation \cite{PSS,FNT}.

In this paper, the effect of the size and shape of the
demagnetized region on the relaxation rate of the total magnetization
is analyzed. In order to demonstrate this effect, we consider the
nonlinear, longitudinal evolution of the total magnetization
following the ultrafast demagnetization for two limiting cases,
namely, for the demagnetized region of cylindrical (CS) and
spherical (SS) symmetries. Then, we compare these results with the
results found within 1d model.

A strongly non-equilibrium state created by the laser pulse plays the role of
the initial condition for LLBar equations \cite{YOI14}. For our case, since the
motion of the wave front is associated with the change in the total
magnetization of the ferromagnet, and the size of a transition region
between demagnetized and non-perturbed regions is much
larger than nanometers, the exchange (nonlocal) term, which retains the total
magnetization of a sample, can be disregarded (compare with
\cite{MentinkPRL12,YOI14}) and the main contribution to the equations
of motion is determined by relativistic processes. Since during the relaxation of magnetization toward an
equilibrium value the effective field is parallel to
the magnetization, only the modulus of the magnetization $M=|\bf
M|$ enters the equation. Adopting the Landau model for the free
energy and introducing dimensionless variables, the nonlinear
diffusion equation (NDE) is derived for the evolution of $M$ \cite{PSS, YOI14}
\begin{equation}
\label{eq2}
\frac{\partial m}{\partial \tau} =\nabla^2 m + m\left(
{1-m^2} \right),
\end{equation}
where $m\left( {\bf {\xi} ,\tau } \right)=M/M_0$ is a  dimensionless
magnetization, $M_0 $ is an equilibrium value
(temperature-dependent) of the magnetization of a bulk material,
$\xi$ is a dimensionless radius in cylindrical or spherical
coordinates measured in units of ${r_0=\sqrt {2A\chi _{\vert \vert }
} }$ (as we consider cylindrically and spherically symmetric
problems, only $\xi$ enters the problem), $A$ is an inhomogeneous
exchange constant, $\chi _{\vert \vert } ={\rd M}/{\rd H}$ is a longitudinal magnetic
susceptibility of a material in the equilibrium state and at zero magnetic
field and $\tau$ is a dimensionless time measured in units of $t_0=2\chi _{\vert \vert
}/\gamma \lambda_\textrm{r} M_\textrm{S}$, $\gamma$ is the gyromagnetic ratio, $\lambda _\textrm{r}$ is a dimensionless relaxation
constant of the relativistic nature, $M_\textrm{S}$ is the saturation magnetization. Simple estimates for nickel show that the value $r_0$ is of the order of a lattice constant, the characteristic time $t_0$
is of the order of a few picoseconds and the characteristic velocity
$r_0/t_0 \approx 0.1~\text{nm/ps} = 100~\text{m/s}$, see for details \cite{PSS}.
Note that the use of the continuum approximation for distributions with
characteristic sizes of the order of the lattice constant does not
lead to qualitative errors, which can be seen from comparison of the
results obtained numerically for discrete models and for their
continual counterpart, see, e.g., \cite{IvKolW96,IvWysin02}.
It is worth noting that this kind of NDE was first studied in the
pioneering work by Kolmogorov, Petrovsky and Piskunov
\cite{KolmPP37} and by Fisher \cite{Fisher37} and stationary
diffusive front propagations into unstable state have been found for
this problem.

After the pulse action, the demagnetized region (spot) is formed in
a sample with a characteristic size of the order of a diameter of
the laser beam. The value of magnetization is reduced inside the
spot, $m=m_0 <1$, and outside the spot the magnet is non-perturbed,
$m=1$. We consider two cases, when the demagnetized region has a
cylindrical and spherical symmetries. To model such a situation, NDE
is numerically solved for the following initial conditions
\begin{equation}
\label{eq3}
m\left( {\xi ,\tau =0} \right)=m_0+\frac{1-m_0}{
1+\mathrm{exp}\left[
{-4{\left( {\xi -R_0 } \right)}/a} \right] }\,,
\end{equation}
where $R_0 $ is the radius of the demagnetized region, a parameter
$a={(1-m_0)\left( \rd m/\rd\xi \right)^{-1}
}|_{m=\left(1+m_0\right)/2}$ describes the characteristic
width of the transition region in the initial conditions. In the
region of the action of the laser pulse $\left( {\xi <R_0 } \right)$,
the magnetization tends to $m_0 $ and outside this region $\left(
{\xi >R_0 } \right)$ $m\left( {\xi ,\tau=0 } \right)$ tends to its
equilibrium value 1.

For a numerical analysis we consider the limiting case, when the
laser pulse is focused to a diffraction limited spot of the order of
1~\textmu{}m \cite{Dvornik13}. The following values of the
parameters (in the dimensionless units) correspond to the following situation:
a radius of the demagnetized region (laser pump) $R_0=1250$, the width of transition region $a=300$
and the radius of the sample $R=1850$. We consider the evolution of the initial state (\ref{eq3}) for the
following minimal values  of magnetization in the demagnetized
region: $m_0=0.9,\;0.5,\;10^{-2},\;10^{-4},\;0.$ The values $m_0
=0.5,\;0.9$ can be realized for a weak intensity of the laser pulse and correspond to the
effective temperature of the spin system lower than the Curie
temperature $T_\textrm{C} $. The values $m_0 =10^{-2},\;10^{-4},\;0$ can be
realized for a high power of the laser pulse.

\section{Cylindrically symmetric initial distribution of magnetization}

\begin{figure}[!b]
\centering
\subfigure[]{\includegraphics[width =  0.38\textwidth]{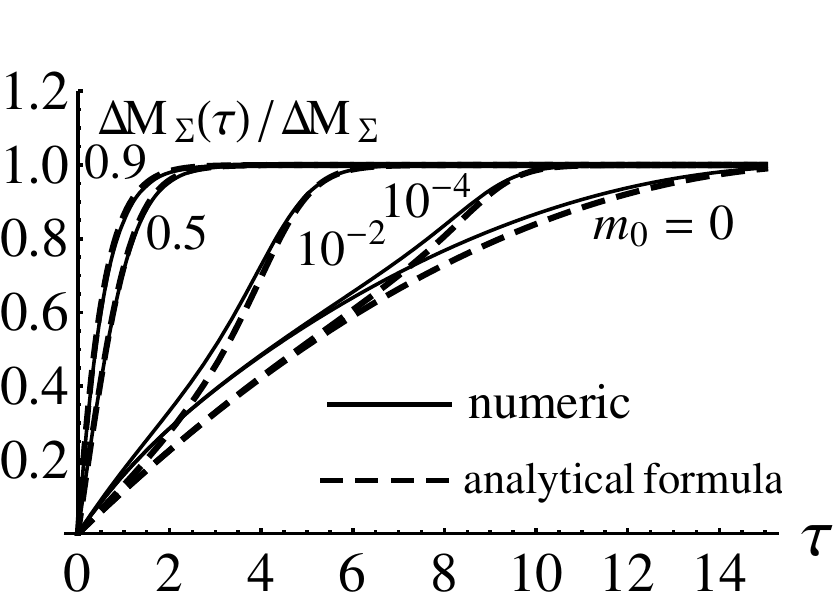}
 \label{11}
 }
 \subfigure[]{\includegraphics[width = 0.60\textwidth]{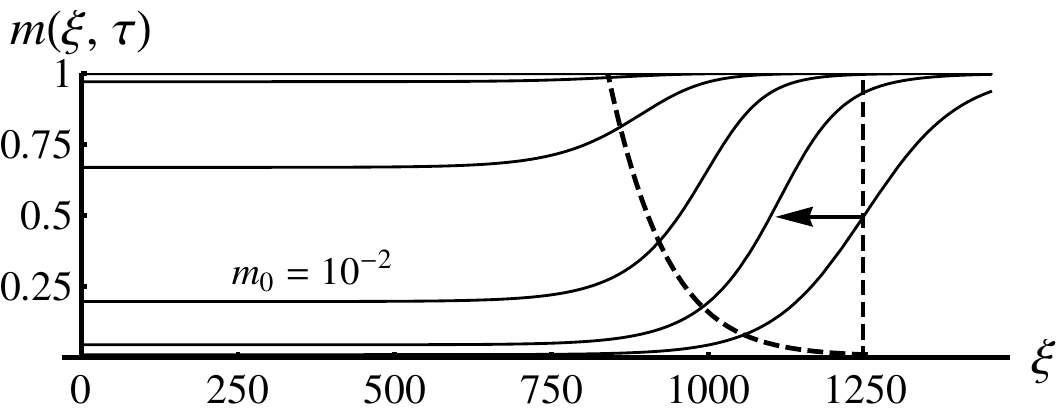}
 \label{22}
 }
\caption{(a) The time evolution of ${\Delta M_\Sigma \left( \tau \right)}/\Delta M_\Sigma$, derived from the numerical solution of NDE for the above chosen numerical parameters, different values of $m_0$ and for CS initial conditions. Full lines represent numerical
calculations and dashed lines are an approximation with the analytical
equation (\ref{eq8}). (b) A corresponding to figure~\ref{fig}~(a) evolution of the magnetization $m(\xi, t)$ for $m_0=10^{-2}$ on a time domain from $\tau =0$ to $\tau =T=15$, taken step-by step after an interval of
time $T/10$. Dashed lines show the contributions of the wave
front and the homogeneous evolution of magnetization. The arrow
indicates a direction of motion of the wave front.} \label{fig}
\end{figure}

Figure~\ref{fig}~(a) presents the time evolution of the relative change of magnetization per unit length of the sample $M_\Sigma \left( \tau \right)$
\begin{equation}
\label{eq4}
\frac{\Delta M_\Sigma \left( \tau \right)}{\Delta M_\Sigma }=\frac{M_\Sigma
\left( \tau \right)-M_\Sigma \left( 0 \right)}{M_\Sigma \left( {+\infty }
\right)-M_\Sigma \left( 0 \right)}\,,
\end{equation}
calculated in the time domain from $\tau =0$ to $\tau =T=15$ (in the
dimensionless units $t_0 )$, for the above chosen parameters and for the initial conditions (\ref{eq3}). Figure~\ref{fig}~(b) demonstrates an example of the corresponding evolution of magnetization $m\left( {\xi ,\tau } \right)$
on the time for $m_0=10^{-2}$.

Figure~\ref{fig}~(a) shows that the fastest regime of relaxation of the total
magnetization to its equilibrium value is realized for $m_0 =0.9$. The
relaxation time of ${\Delta M_\Sigma \left( \tau \right)}/\Delta M_\Sigma$, as for 1d model \cite{FNT}, decays with
a decrease of $m_0 $ and takes its minimum value at $m_0 =0$. However, the dependence of ${\Delta M_\Sigma \left( \tau \right)}/\Delta M_\Sigma$ on time is not linear for $m_0 =0$ in CS case. The analysis of the evolution of $m\left( {\xi ,\tau } \right)$ [figure~\ref{fig}~(b)] demonstrates that similarly to 1d model \cite{PSS, FNT}, the evolution of
the total magnetization to its equilibrium value is determined by
two effects: the homogeneous evolution of the magnetization and the motion
of the wave front within the demagnetized region.

Let us derive an approximate analytical formula for the evolution of
$M_\Sigma \left( \tau \right)$ for CS case, which takes into account both the
motion of the wave front and the homogeneous evolution of the
magnetization and estimate the contribution of the wave front
compared to 1d model. Following \cite{FNT}, the evolution of the wave
front can be cast in the form
\begin{equation}
\label{eq6}
m\left( {\xi ,\tau } \right)\approx m_0 \left( \tau \right)+\left[ {1-m_0
\left( \tau \right)} \right]m_\textrm{f} \left( {\xi -\xi _0 -V\tau,\xi} \right),
\end{equation}
where $m_0 \left( \tau \right)$ does not depend on the coordinate
and describes the homogeneous evolution of magnetization within
the demagnetized region. The second term in the right-hand side
of (\ref{eq6}) corresponds to the wave front, which moves within the demagnetized region with an amplitude $1-m_0 \left( \tau
\right)$ and the velocity $V$. The function $m_\textrm{f} $ describes the
shape of this front; the value of $m_\textrm{f} $ tends to $1$ if one moves
outside the demagnetized region from the wave front and $m_\textrm{f} $ tends
to $0$ within the demagnetized region. Both $V$ and the shape of
the wave front depend on the distance from the center of the
demagnetized region $\xi$. However, this problem can be simplified
by the following observation: figure~\ref{fig}~(b) shows that  the contribution of
the wave front is essential only at the initial stage of the
evolution. This means that: (i) one can neglect the term $\left(
{1/\xi}\right){\partial m}/\partial \xi \propto 1/{\left( {a\cdot
\xi } \right)}\ll 1/{a^2}$ and (\ref{eq2}) transforms to 1d NDE;
(ii) we can suppose that within the demagnetized region $m\left(
{\xi ,\tau } \right)\simeq m_0 \left( \tau\right)\ll 1$ and the
terms of the order of $m_0 \left( \tau \right)\ll 1$ can be
neglected in this approximation. Then, the shape $m_\textrm{f}$ coincides
with the  shape of the wave front at $m_0 =0$ in 1d NDE and the
velocity of the wave front $V$ formed from the initial conditions
can be found by equating the asymptotic of (\ref{eq3}) within
the demagnetized region and that of the wave front: $V =a/4+4/a$
\cite{Kametaka}. For the chosen numerical parameters $V \approx
75.01$.
Neglecting dispersion in (\ref{eq2}) and integrating this
equation, the dependence of the magnetization on time $m_0(\tau)$
within the demagnetized region can be present as follows \cite{YOI14}:
\begin{equation}
\label{eq7}
m_0(\tau)=\frac{m_0}{\sqrt{m_0^2 +\left( {1-m_0^2 }
\right)\exp \left( {-2\tau } \right)}}\,.
\end{equation}
Thus, at $m_0 \left( \tau \right)\ll 1$ the homogeneous evolution of
magnetization leads to a decrease of the amplitude of the wave
front $\left[ {1-m_0 \left( \tau \right)} \right]$, and the motion
of the wave front leads a decrease of the characteristic radius
of the demagnetized region, which is equal to $\left( {R_0 -V\tau }
\right)$.

Integrating (\ref{eq6}) over the unit length of the cylinder, the estimating formula for
the evolution of $M_\Sigma \left( \tau \right)$ is derived
\begin{equation}
\label{eq8}
M_\Sigma \left( \tau \right)\approx M_\Sigma \left( 0 \right)+\pi R_0^2 m_0
\left( \tau \right)+\pi \left[ {1-m_0 \left( \tau \right)} \right]\left(
{2R_0 -V\tau } \right)V\tau ,
\end{equation}
where $m_0 \left( \tau \right)$ is defined by (\ref{eq7}). Equation (\ref{eq8}) takes into account both
the homogeneous evolution of magnetization and the motion of the wave
front.

Figure~\ref{fig}~(a) shows a good agreement between the analytical equation (\ref{eq8}) and the numerical
results even in the regime, at which the magnetization within the
demagnetized region $m_0 \left( \tau \right)$ is not small, when both the
shape and the velocity of the wave front depend on $m_0 \left( \tau
\right)$. The reason here is that for this regime the contribution of the wave front becomes small.

Let us estimate the relative contribution of the wave front and the
homogeneous evolution of magnetization for different $m_0$ for CS case. We present
\begin{equation}
\label{eq9}
\frac{\rd M_\Sigma \left( \tau \right)}{\rd\tau}=
\frac{\rd M_\Sigma ^\textrm{wave} \left(
\tau \right)}{\rd\tau}+\frac{\rd M_\Sigma^\textrm{flat} \left( \tau \right)}{\rd\tau}\,,
\end{equation}
where $M_\Sigma^\textrm{wave}\left( \tau \right)$ and $M_\Sigma
^\textrm{flat} \left( \tau \right)$ are contributions of the wave front and the homogeneous evolution, respectively
[these contributions are schematically depicted in figure~\ref{fig}~(b) by dashed lines].
Differentiating (\ref{eq8}) with respect to time and comparing the
result with (\ref{eq9}), one can see that
\begin{equation}
\label{eq10} \frac{\rd M_\Sigma ^\textrm{wave} }{\rd \tau }\simeq 2\pi \left(
{R_0 -V\tau} \right)\left[{ 1-m_0 \left( \tau \right) } \right]V, \qquad
 \frac{\rd M_\Sigma ^\textrm{flat} }{\rd\tau }\simeq \pi \left( {R_0
-V\tau } \right)^2\left[ {m_0 \left( \tau \right)-m_0 \left( \tau
\right)^3} \right].
\end{equation}
Here, $\left[ {1-m_0 \left( \tau \right)} \right]$ is the amplitude
of the wave front, $\left( {R_0 -V\tau } \right)$ is the
characteristic radius of the demagnetized region at the moment $\tau
$.

Thus, ${\rd M_\Sigma^\textrm{wave}\left( \tau \right)}/\rd\tau$ takes its
maximum value at the initial stage of the evolution. However,
further, the the relaxation rate caused by the wave front decreases
with a decrease of both the amplitude and the length of the wave front.

\begin{table}[!b]
\begin{center}
\begin{tabular}{|p{96pt}|p{35pt}|p{35pt}|p{35pt}|p{35pt}|}
\hline\hline
$m_0 $&
$0.5$&
$10^{-2}$&
$10^{-4}$&
$0$ \\
\hline
${M_\Sigma ^\textrm{wave}}/{\Delta M_\Sigma } (\text{1d})$&
$0.05$&
$0.24$&
$0.51$&
$1$ \\
\hline
${M_\Sigma ^\textrm{wave}}/{\Delta M_\Sigma } (\text{CS})$&
$0.09$&
0.41&
$0.75$&
$1$ \\
\hline
${M_\Sigma ^\textrm{wave}}/{\Delta M_\Sigma } (\text{SS})$&
$0.13$&
0.53&
$0.85$&
$1$ \\
\hline\hline
\end{tabular}
\end{center}
\caption{The relative contribution of the wave front for 1d, CS and
SS cases for the above chosen values of the parameters.\label{tab1}}
\end{table}

The relative contribution of the wave front can be defined as ${M_\Sigma
^\textrm{wave} \left( \tau \right)}/{\Delta M_\Sigma }$ at $\tau \to +\infty $, where $M_\Sigma ^\textrm{wave} \left(
\tau \right)$ can be found by integrating (\ref{eq10}).
Table~\ref{tab1} presents ${M_\Sigma^\textrm{wave} \left( \tau \right)}/{\Delta M_\Sigma }$ for different $m_0 $, for 1d model \cite{FNT}, CS and SS cases.
Table~\ref{tab1} demonstrates that the contribution of the wave front for CS is
stronger than for 1d model (almost two times stronger for not small $m_0$).
To understand this observation one should note that the wave front gives the main contribution at the initial stage. Then, estimating ${\rd M_\Sigma ^\textrm{wave} }/{(\Delta M_\Sigma \rd\tau)}$
at $\tau =0$ (initial stage), we see that
this value is two times larger in CS case than that found within 1d model. With a decrease of $m_0 $, ${M_\Sigma ^\textrm{wave}}/ {\Delta M_\Sigma }$ increases and takes its maximum value $1$ for $m_0 =0$.

To estimate the value of magnetization in the demagnetized region
$m_\textrm{eq}^\textrm{CS}$ (for CS case), for which the relaxation rates for the wave front and the
homogeneous evolution become comparable, we equate
${\rd M_\Sigma^\textrm{wave}}/{\rd\tau } {=\rd M_\Sigma ^\textrm{flat}}/{\rd\tau }$.  This is a nonlinear, transcendental equation. However, a simple estimating formula can be derived by noting that the wave front gives the main contribution at the initial stage, that is $m_\textrm{eq}^\textrm{CS}\ll 1$. Then, neglecting the terms like $(m_\textrm{eq}^\textrm{CS})^2$, we get
\begin{equation}
\label{eqa}
m_\textrm{eq}^\textrm{CS} \simeq 2V/R_0\,.
\end{equation}

Thus, if the initial value of magnetization in the demagnetized region for CS situation $m_0\ll m_\textrm{eq}^\textrm{CS},$  the wave front gives a substantial contribution to the relaxation of the total magnetization. Note, that $m_\textrm{eq}^\textrm{CS}$ is two times larger than for 1d model $m_\textrm{eq}^\textrm{1d} \simeq {V}/{R_0}$ \cite{FNT}.

\section{Spherically symmetric initial distribution of magnetization}

The evolution of the wave front for SS case can be also presented in
the form (\ref{eq6}). Using the assumptions similar to the CS case and
integrating (\ref{eq6}) over the volume of the sample, the
estimating formulae are derived
\begin{equation}
\label{eq12} M_\Sigma \left( \tau \right)\approx M_\Sigma \left( 0
\right)+\frac{4\pi}{3} R_0^3 m_0 \left( \tau \right)+4\pi
V\tau\left[ {1-m_0 \left( \tau \right)} \right]\left( {R_0^2 - R_0
V\tau+\frac{V^2\tau^2  }{3} } \right) ,
\end{equation}
where $m_0 \left( \tau \right)$ is also defined by (\ref{eq7}), $M_\Sigma \left( \tau \right)$ is the total magnetization of the sample and
\begin{equation}
\label{eq13} \frac{\rd M_\Sigma ^\textrm{wave} }{\rd\tau }\simeq 4\pi \left(R_0
-V\tau \right)^2\left[{ 1-m_0 \left( \tau \right) } \right]V, \qquad
\frac{\rd M_\Sigma ^\textrm{flat} }{\rd\tau }\simeq \frac{4\pi }{3 } \left( {R_0 -V\tau }
\right)^3\left[ {m_0 \left( \tau
\right)-m_0 \left( \tau
\right)^3} \right].
\end{equation}
These formulae well describe the results of the numerical
calculations. Table~\ref{tab1} demonstrates that the contribution of the wave
front for SS case is stronger than for CS case (almost $1.5$ times for not small $m_0$). Estimating ${\rd M_\Sigma ^\textrm{wave} }/{(\Delta M_\Sigma \rd\tau)}$ at the
initial stage, we see that ${\rd M_\Sigma ^\textrm{wave} }/{(\Delta M_\Sigma
\rd\tau)}$ at $\tau =0$ for SS case is $1.5$ times larger than in the CS case, thus giving the explanation of the numerical data.
Comparing different contributions in \ref{eq13}, we get
\begin{equation}
\label{eq15}
m_\textrm{eq}^\textrm{SS} \simeq 3V/R_0.
\end{equation}
So, $m_\textrm{eq}^\textrm{SS}$ is $1.5$ times larger than for the CS case.

\section{Conclusion}
The relaxation of highly nonequilibrium, nonuniform states created by the femtosecond laser pulse in the relativistic relaxation approximation is governed by two scenarios: the motion of the wave front within the demagnetized region and the homogeneous evolution of magnetization within the demagnetized region. The relativistic relaxation approximation is valid for widths of the transition regions between the demagnetized and non-perturbed regions exceeding a few dozens nanometers (compare with \cite{YOI14}). The contribution of the wave front for the SS shape of the demagnetized region is up to $3$ times and for CS case
is up to $2$ times stronger than for 1d model \cite{FNT}; thus, the motion of the wave front within the demagnetized region leads to the dependence of the relaxation rate on the shape of the
demagnetized region. For a sufficiently large diameter of the demagnetized region (laser spot), the homogeneous evolution dominates and the evolution of the total magnetization is the same
for 1d model, CS and SS cases. However, with a decrease of the diameter of the demagnetized region and with an increase of the power of the laser impulse, the relative contribution of the wave front increases and for $m_0<(m_\textrm{eq}^\textrm{CS},m_\textrm{eq}^\textrm{SS})$ the wave front can significantly enhance the relaxation and, consequently, the shape dependent effects can also be  observed. This situation could be realized after demagnetization by a tightly focused femtosecond laser impulse.

\section*{Acknowledgements}
I thank prof. B.A.~Ivanov for the help in the formulation of the problem and for fruitful discussions.

\vspace{-3mm}

\ukrainianpart

\title{Поздовжня еволюція намагніченості феромагнетика після надшвидкого розмагнічування: роль розміру та форми розмагніченої облаcті}
\author{І.О. Ястремський}
\address{
Київський національний університет імені Тараса Шевченка, \\ вул. Володимирська, 64,  01601 Київ, Україна
}

\makeukrtitle

\begin{abstract}
\tolerance=3000%
Продемонстровано залежність швидкості релаксації поздовжньої еволюції повної намагніченості феромагнетика після надшвидкого розмагнічування від форми розмагніченої області. Ця залежність обумовлена рухом хвильового фронту вглиб розмагніченої області. Внесок хвильового фронту для сферично-симетричної форми розмагніченої області до 3 раз та для циліндрично-симетричної області до 2 раз більший, ніж для одновимірного випадку. Цей ефект може спостерігатись після розмагнічування сильно сфокусованим лазерним імпульсом.

\end{abstract}

\end{document}